\documentclass[conference]{IEEEtran}

\usepackage{graphicx}
\usepackage{amsmath}
\usepackage{url}
\usepackage{cite}
\usepackage{hyperref}

\hypersetup{
    colorlinks=true,
    linkcolor=blue,
    citecolor=blue,
    urlcolor=blue
}

\usepackage{tikz}
\usetikzlibrary{positioning, shapes.geometric, arrows.meta, calc, quotes, arrows, decorations.markings, shadows.blur}
\usepackage[T1]{fontenc}
\usepackage{xcolor}
\usepackage[font=footnotesize,labelfont=bf]{caption}

\title{TRACE: AI-Assisted Assessment of Collaborative Projects in Computer Science Education}

\author{
\IEEEauthorblockN{Songmei Yu}
\IEEEauthorblockA{
\textit{Felician University} \\
yus@felician.edu \\
}
\and
\IEEEauthorblockN{Andrew Zagula}
\IEEEauthorblockA{
\textit{Bridgewater-Raritan High School} \\
andrewzagula800@gmail.com \\
}
}

\begin{document}

\maketitle

\begin{abstract}
Collaborative group projects are integral to computer science education, fostering teamwork, problem-solving, and industry-relevant skills. However, assessing individual contributions within group settings remains challenging. Traditional approaches, including equal grade distribution and subjective peer evaluations, often lack fairness, objectivity, and scalability, particularly in large classrooms. We propose TRACE, a semi-automated AI-assisted framework for assessing collaborative software projects that evaluates both project quality and individual contributions using repository mining, communication analytics, and AI-assisted analytics. A pilot deployment in a software engineering course demonstrated high alignment with instructor assessments, increased student satisfaction, and reduced instructor grading effort. The results suggest that AI-assisted analytics can improve the transparency and scalability of collaborative project assessment in computer science education.
\end{abstract}

\section{Introduction}

Group projects are foundational to computer science curricula, providing students with hands-on experience in software engineering practices such as Agile methodologies, version control, collaboration, and iterative testing. These collaborative activities mirror professional software development environments and help prepare students for real-world workflows. However, assigning fair individual grades remains challenging, particularly when team sizes are large or contributions are unevenly distributed.

Instructors often lack granular visibility into team dynamics, leading to assessments based on incomplete information. Peer evaluations, while useful, may be influenced by bias, social dynamics, or non-academic considerations. Manual oversight becomes difficult to maintain at scale, creating a need for automated tools that enable accurate and transparent assessment.

This work introduces TRACE, an AI-assisted grading framework for collaborative software projects. TRACE (1) evaluates overall project quality using standardized and customizable metrics, (2) quantifies individual contributions using objective signals such as version control activity and communication logs, (3) generates fair and explainable individual grades, (4) reduces instructor workload while maintaining instructor oversight, and (5) improves transparency in collaborative project assessment.

\section{Related Work}

Traditional approaches to grading collaborative projects include equal grade distribution, peer evaluation tools, and instructor observation. Equal grading assigns the same score to all team members regardless of individual contribution. Peer evaluation systems, such as CATME and SPARKPLUS, attempt to assess team member performance from both social and task-based perspectives. However, these systems are susceptible to bias, collusion, and peer pressure \cite{ohland2012catme,kaufman2000peer}. In addition, peer reviews may fail to accurately reflect technical contributions, particularly in large or distributed teams \cite{hamer2005tools}.

Instructor observation seeks to estimate individual effort through project meetings, code reviews, or other limited interactions. However, this approach is time-consuming, difficult to scale, and often inconsistent, particularly in large classes or remote settings \cite{freeman2011destructive,gehringer2001peerreview}.

To address these limitations, recent research has explored analyzing software repositories to estimate individual contributions. Metrics such as commit counts, lines of code changed, file ownership, and authorship of specific code segments provide quantifiable proxies for student effort \cite{bird2009ownership,spinellis2012platform,vasilescu2015blogs,rigby2011broadcast}. However, commit frequency alone does not necessarily indicate meaningful contribution, as students may generate frequent but trivial changes (e.g., formatting updates or whitespace edits). Tools such as GitGrade, Codequiry, and Gitorium extract basic metrics from version control systems but often lack integrated frameworks that evaluate contribution quality or project context.

More advanced approaches apply Natural Language Processing (NLP) to analyze issue trackers, pull request discussions, and documentation comments \cite{tsay2014github,wang2016interactions,mcdonald1998expertise}. These methods provide a broader view of collaboration by capturing communication patterns, leadership roles, and task complexity in addition to code contributions.

The broader field of Artificial Intelligence in Education (AIEd) applies machine learning and educational data mining to support adaptive instruction and feedback \cite{siemens2013learning}. However, automated grading systems are frequently criticized for limited transparency, reduced adaptability, and reliance on historical patterns that may disadvantage students with atypical learning styles.

Finally, ethical considerations, including algorithmic fairness, data privacy, surveillance, and student consent, are critical \cite{williamson2017bigdata}. These must be carefully addressed before integrating AI into high-stakes assessments, such as course grades.

\section{System Design and Architecture}

TRACE consists of three core modules: the Project Quality Assessment Module (PQAM), the Individual Contribution Analyzer (ICA), and the Grading Engine (GE). The following sections describe each component and its interactions within the system.

\subsection{Project Quality Assessment Module (PQAM)}

The Project Quality Assessment Module (PQAM) (Fig.~\ref{fig:pqam}) evaluates the technical and qualitative integrity of a group project. It combines static analysis, repository analytics, and AI-assisted evaluation to determine whether a project meets academic and industry standards. PQAM consists of five submodules: Code Quality, Testing Coverage, Documentation, Functionality, and Usability.

\subsubsection{Code Quality}

The Code Quality submodule uses static analysis to examine the source code without executing it. It detects structural flaws, maintainability issues, and stylistic violations in the codebase.

First, the system computes complexity metrics, including (1) Cyclomatic Complexity, which measures the number of independent execution paths in the code, and (2) Halstead Metrics, which estimate computational complexity based on operators and operands. These metrics provide insight into code readability, maintainability, and potential error-proneness.

Second, the system evaluates style adherence to verify compliance with language-specific standards. For example, Python code can be analyzed using tools such as \texttt{flake8}, \texttt{pylint}, and \texttt{black} to enforce PEP8 guidelines, while JavaScript projects may use \texttt{ESLint} and \texttt{Prettier} to ensure consistency with established coding conventions.

Third, the system performs code duplication detection using tools such as SonarQube, PMD, or jscpd to identify copy-paste patterns that indicate poor abstraction or limited code reuse.

\subsubsection{Testing Coverage}

The Testing Coverage submodule evaluates how thoroughly tests cover the codebase. Coverage tools such as \texttt{pytest-cov} (Python), JaCoCo or JUnit integrations (Java), and Istanbul or Jest (JavaScript) generate reports that measure branch, path, and function-level coverage.

Beyond simple coverage percentages, the system can analyze whether edge cases, exception handling, and input validation are adequately tested, providing a deeper view of test robustness.

\subsubsection{Documentation}

The Documentation submodule evaluates the clarity, completeness, and technical quality of project documentation using Natural Language Processing (NLP) techniques. Models such as BERT, spaCy, or GPT-based classifiers can assess grammar, structure, and coherence in README files and project manuals.

Markdown parsing and LLM-based validation are used to verify the presence of essential components, including installation instructions, usage documentation, architectural descriptions, contributor guidelines, and licensing information. Inline comments and docstrings can be analyzed with tools like Doxygen or through semantic similarity analysis to determine whether comments meaningfully describe the underlying code.

Notebook-based documentation or project wikis may also be evaluated to assess clarity of explanation and technical depth.

\subsubsection{Functionality}

The Functionality submodule evaluates whether the system operates as intended. Automated test execution runs predefined or student-provided unit and integration tests within sandboxed environments (e.g., Docker containers). The system compares expected and actual outputs and flags discrepancies.

Continuous integration signals, such as passing builds from GitHub Actions, Travis CI, or GitLab CI/CD pipelines, can also be incorporated. In addition, feature validation may be performed through log analysis or semantic parsing of project requirements to verify that implemented functionality matches expected deliverables.

\subsubsection{Usability}

The Usability submodule evaluates the accessibility and usability of systems that include graphical or web-based interfaces. Heuristic evaluation methods based on established usability principles (e.g., Nielsen’s heuristics) can identify issues such as inconsistent navigation, missing feedback, or poor error messaging.

Automated UI testing frameworks such as Selenium, Cypress, or Puppeteer enable cross-browser interaction testing. Accessibility tools, including Lighthouse, Axe Monitor, and Pa11y, can verify compliance with the Web Content Accessibility Guidelines (WCAG), including checks for color contrast, screen-reader compatibility, and keyboard navigation. Responsiveness testing further evaluates whether interfaces function correctly across different screen resolutions and device types.

\subsubsection{Scoring} 

PQAM aggregates outputs from the five submodules into a unified Project Quality Score (PQS). By default, each component is weighted equally, though instructors may adjust weights to align with course objectives or grading rubrics.

In addition to producing a numerical score, PQAM generates a project-quality dashboard that summarizes strengths and weaknesses across dimensions such as code maintainability, test robustness, documentation completeness, and usability.

\begin{figure}[t]
\centering
\begin{tikzpicture}[
  font=\sffamily\footnotesize,
  node distance=4mm and 8mm,
  labelbox/.style={
    draw,
    fill=blue!20,
    text width=10mm,
    align=center,
    minimum height=3.5cm,
    rounded corners
  },
  subbox/.style={
    draw,
    fill=green!15,
    text width=3cm,
    align=center,
    minimum height=6.5mm,
    rounded corners
  },
  connect/.style={line width=0.6pt, draw=black, -{Latex[length=2mm]}}
]

\node[subbox] (a) {Code Quality};
\node[subbox, below=of a] (b) {Testing Coverage};
\node[subbox, below=of b] (c) {Documentation};
\node[subbox, below=of c] (d) {Functionality};
\node[subbox, below=of d] (e) {Usability};

\node[labelbox, left=1.5cm of c] (label) {\rotatebox{90}{Project Quality Score}};

\coordinate (merge) at ($(label.east)+(7.5mm,0)$);

\foreach \x in {a,b,c,d,e}
  \draw[connect] (\x.west) -| (merge) -- (label.east);

\end{tikzpicture}
\vspace{3mm}
\caption{Structure of the Project Quality Assessment Module (PQAM).}
\label{fig:pqam}
\end{figure}

\subsection{Individual Contribution Analyzer (ICA)}

The Individual Contribution Analyzer (ICA) (Fig.~\ref{fig:ica}) evaluates each student’s involvement in a collaborative software project. Traditional group assessments often obscure individual effort, potentially leading to unfair grading. ICA addresses this limitation by analyzing multiple signals from the development lifecycle to quantify individual contributions. ICA consists of four submodules: Commit Analysis, Code Ownership, Issue Tracker, and Code Review.

\subsubsection{Commit Analysis} 

Commit Analysis evaluates version control activity, focusing on meaningful code contributions rather than superficial edits. First, trivial changes are filtered using rule-based methods that combine regular expressions and diff analysis to detect non-substantive modifications such as whitespace changes, minor comment edits, or variable renaming. This prevents artificial inflation of contributions through excessive or low-impact commits.

Second, temporal and behavioral patterns are analyzed, including commit frequency, time of day, and burst activity that may indicate rushed development. Density-based clustering algorithms such as DBSCAN can distinguish sustained development activity from last-minute contributions.

Third, quantitative metrics are extracted from commit diffs, including lines added, lines deleted, and net code change. Additional weight may be assigned to commits involving test cases, documentation, or refactoring tasks.

\subsubsection{Code Ownership} 

The Code Ownership component attributes authored code to individual contributors using the \textit{git blame} command. Each line in the final submission is mapped to its author via Git history, and edited lines are tracked to identify both the original authors and subsequent contributors.

Ownership credit is primarily assigned to contributors who introduce or substantially modify code logic, while partial credit is given for minor edits, such as formatting changes. Ownership is aggregated at the function, class, or module level to identify contributors responsible for critical components of the system (e.g., database layers or frontend interfaces).

\subsubsection{Issue Tracker}

Issue Tracker captures non-code contributions that are often overlooked in traditional grading. The system records who creates issues (e.g., bug reports or feature requests), responds to them, resolves them, or references them in commits.

Integration with platforms such as GitHub Issues, GitLab, or Jira enables the extraction of structured collaboration data. Contribution types can be inferred from issue labels (e.g., documentation, performance, bug fix, enhancement) or through NLP analysis of issue descriptions and discussion threads. Engagement metrics include the number of issues created or resolved, as well as responsiveness to team discussions.

\subsubsection{Code Review}

Code Review measures participation in peer code review, an essential practice for improving software quality and collaborative learning. ICA evaluates both the frequency and quality of review activity.

Review frequency is measured by tracking the number of pull request reviews and comments each student submits. NLP-based depth analysis uses models such as BERT or RoBERTa to determine whether reviews identify logical issues, suggest improvements, or reference coding standards. Superficial or generic comments receive lower weight. Sentiment and tone analysis further evaluates whether feedback is constructive, neutral, or overly critical, promoting professional collaboration within teams.

\subsubsection{Scoring}

Finally, ICA aggregates outputs from the four submodules into a unified Normalized Contribution Score (NCS\textit{)} for each student. As in PQAM, each component is weighted equally by default, although instructors may adjust weights to emphasize particular types of contributions.

In addition to producing a numeric score, ICA can generate a contribution dashboard displaying commit timelines, ownership distributions, and participation heatmaps. These visualizations help instructors quickly identify imbalances in team participation.

\begin{figure*}[t]
\centering
\begin{tikzpicture}[
  font=\sffamily\footnotesize,
  node distance=4mm and 12mm,
  box/.style={
    draw,
    fill=gray!15,
    text width=3.4cm,
    align=center,
    minimum height=13mm,
    rounded corners
  },
  subbox/.style={
    draw,
    fill=white,
    text width=3.0cm,
    align=center,
    minimum height=7mm,
    rounded corners
  },
  outputbox/.style={
    draw,
    fill=blue!50!cyan!70,
    text=white,
    text width=3.4cm,
    align=center,
    minimum height=10mm,
    rounded corners
  },
  connect/.style={
    line width=0.6pt,
    draw=black,
    postaction={decorate},
    decoration={markings, mark=at position 0.5 with {\arrow{stealth}}}
  }
]

\node[subbox] (a) {Commit Analysis};
\node[subbox, below=of a] (b) {Code Ownership};
\node[subbox, below=of b] (c) {Issue Tracker};
\node[subbox, below=of c] (d) {Code Review};

\node[coordinate] (midpoint) at ($(a.north)!0.5!(d.south)$) {};

\node[box, left=3.0cm of midpoint] (input) {Repository and\\Communication Artifacts};
\node[outputbox, right=3.2cm of midpoint] (output) {Normalized\\Contribution Score};

\foreach \x in {a,b,c,d}
  \draw[connect] (input.east) -- (\x.west);

\foreach \x in {a,b,c,d}
  \draw[connect] (\x.east) -- (output.west);

\end{tikzpicture}
\vspace{4mm}
\caption{Structure of the Individual Contribution Analyzer (ICA).}
\label{fig:ica}
\end{figure*}

\subsection{Grading Engine (GE)}

The Grading Engine (GE) is the final layer of the AI-assisted assessment pipeline. It performs weighted aggregation to generate final individual grades from group-level project outcomes and individual contribution signals. GE computes each student’s final score by combining the Project Quality Score (PQS) produced by PQAM, which represents the overall technical quality of the group’s work, and the Normalized Contribution Score (NCS) produced by ICA, which reflects each student’s relative contribution within the team. Both PQS and NCS are transformed to a common 0-100 scale before aggregation. The default formula is:
\begin{equation}
Score = (PQS \cdot 0.6) + (NCS \cdot 0.4)
\end{equation}

This formula is configurable. Instructors may adjust the weights to emphasize team output (e.g., 80/20) or individual effort (e.g., 50/50), depending on pedagogical objectives. Grades are generated automatically but are flagged for manual review when anomalies (e.g., contribution outliers) are detected.

To support fair grading across teams with different sizes and collaboration patterns, NCS scores may be normalized. For example, Z-score normalization or Min-Max scaling can scale each student’s raw contribution score within the team, enabling meaningful comparison and reducing the influence of team composition. Extremely low contributions (e.g., <10\%) may trigger floor limits, while unusually high contributions may be capped to reduce grade inflation and promote equity. Instructors may also assign bonuses for exemplary contributions or penalties for under-participation based on team evaluations or reflective reports.

GE detects irregular or potentially unfair patterns and flags them for human review. Statistical methods such as the interquartile range and standard deviation identify contributors whose participation differs significantly from that of their peers, as well as teams with unusually high PQS values but low NCS engagement. Additional inconsistencies may be detected, including mismatches between contribution levels and code authorship (e.g., frequent commits with minimal authored code), or uneven participation in code reviews and issue resolution. Flagged submissions are queued in an instructor dashboard that presents visualizations of contribution distributions, suggested rubric-based adjustments, and space for instructor notes or overrides.

A central objective of GE is to improve transparency and fairness in grading. The system generates detailed reports for students that include PQS, NCS, and the final score. Instructors receive visual analytics such as radar charts and contribution timelines. GE can also generate feedback prompts that summarize performance patterns, for example: “Code coverage and documentation were strong, but code review participation was limited,” or “Project quality was high, but individual contribution was significantly below the team average.” All grading decisions and manual overrides are logged to support transparency during grade reviews or disputes.

The Grading Engine (GE) serves as the interface between AI-driven analysis and final grading outcomes. By combining group-level quality metrics with individual contribution measures and anomaly detection, GE supports fair, evidence-based grading while preserving instructor oversight and judgment.

\section{System Implementation and Analysis}

To evaluate the effectiveness, fairness, and scalability of TRACE, a pilot deployment was conducted during a software engineering course at a university in Spring 2025. The course included 20 students divided into five teams, each completing collaborative projects over an 8-week period. Projects followed Agile methodologies and required the use of GitHub for version control, issue tracking, and code submissions.

TRACE was implemented in Python. The backend uses Flask to provide RESTful APIs and PostgreSQL for relational data storage, while frontend dashboards built with React.js allow instructors and students to view project metrics, grades, and feedback. The analytics pipeline integrates several libraries and analysis tools. Static code analysis is performed using \texttt{Radon} and \texttt{Pylint}, which compute metrics such as cyclomatic complexity and style compliance. Repository mining is conducted through \texttt{GitPython}, enabling extraction of commit history and file authorship information. Natural language processing components rely on \texttt{spaCy} and \texttt{NLTK} to analyze README files, code comments, and issue tracker discussions, while \texttt{scikit-learn} and \texttt{TensorFlow} support machine learning models used in the analytics pipeline.

Although the pilot deployment focused on repository mining, static analysis, authorship tracking, and communication analytics, the TRACE architecture is designed to support additional evaluation modules. These include documentation analysis, usability and accessibility testing, and automated analysis of code review interactions, allowing the system to extend beyond the capabilities used in the initial deployment.

\subsection{Workflow Overview}

TRACE follows a modular five-step pipeline.

\subsubsection{Project Submission}

Students submitted final codebases and documentation through GitHub Classroom, ensuring version control and a consistent repository structure.

\subsubsection{Data Extraction}

Repositories were programmatically mined using the GitHub REST API and GraphQL endpoints to extract commit histories and diffs, issue tracker logs, pull requests, code reviews, README files, wikis, and inline comments.

\subsubsection{Model Evaluation}

PQAM evaluated group-level project quality using static analysis, test coverage metrics, documentation assessment, and usability checks. ICA analyzed Git metadata and communication records to infer individual contributions using attribution modeling, NLP, and statistical analysis.

\subsubsection{Grade Calculation}

The Grading Engine (GE) combined PQS and NCS using a configurable formula. Anomaly detection algorithms flagged unusually low or high contributions for manual review.

\subsubsection{Dashboard Review}

Instructors accessed a secure web portal to review performance breakdowns and individual contributions, adjust or approve final grades, and export results to the LMS.

Fig.~\ref{fig:workflow} illustrates the system workflow from submission to review, highlighting the automated sections of the pipeline and points where instructor oversight is applied.

\begin{figure}[htbp]
\centering
\resizebox{0.98\columnwidth}{!}{
\begin{tikzpicture}[
  >=Stealth,
  arrow/.style={->, thick, black},
  autobox/.style={
    rectangle, rounded corners=4pt,
    draw=blue!50!black, fill=blue!5,
    text width=5cm, minimum height=1.2cm,
    align=center, font=\sffamily\small
  },
  decision/.style={
    diamond, aspect=2.4,
    draw=orange!60!black, fill=orange!10,
    text width=2.4cm, inner sep=1pt,
    align=center, font=\sffamily\small
  },
  humanbox/.style={
    rectangle, rounded corners=4pt,
    draw=gray!70!black, fill=gray!10,
    text width=4.3cm, minimum height=1.4cm,
    align=center, font=\sffamily\small
  },
  io/.style={
    trapezium, trapezium left angle=70, trapezium right angle=110,
    draw=blue!60!black, fill=blue!15,
    text width=7.2cm, minimum height=1.5cm,
    align=center, font=\sffamily\small
  }
]

\node[autobox] (s1) {Step 1: Project Submission\\\footnotesize Code + Documentation};
\node[autobox, below=of s1] (s2) {Step 2: Data Extraction\\\footnotesize Commits, PRs, issues, docs};
\node[autobox, below=of s2] (s3) {Step 3: Model Evaluation\\\footnotesize PQAM + ICA};
\node[autobox, below=of s3] (s4) {Step 4: Grading Engine\\\footnotesize Final score};
\node[decision, below=of s4] (d) {Anomaly?};

\node[autobox, below left=1.2cm and 0.9cm of d] (man)
  {Instructor Adjustment};

\node[autobox, below right=1.2cm and 0.9cm of d] (s5)
  {Step 5: Dashboard\\\footnotesize LMS export};

\draw[arrow] (s1) -- (s2);
\draw[arrow] (s2) -- (s3);
\draw[arrow] (s3) -- (s4);
\draw[arrow] (s4) -- (d);

\draw[arrow] (d) -- node[above left, font=\sffamily\scriptsize]{Yes} (man);
\draw[arrow] (d) -- node[above right, font=\sffamily\scriptsize]{No} (s5);
\draw[arrow] (man.east) -- (s5.west);

\end{tikzpicture}
}
\vspace{0mm}
\caption{Full TRACE Workflow.}
\label{fig:workflow}
\end{figure}

\subsection{Metrics and Evaluation Methods}

To evaluate system validity and instructional impact, three assessment methods were employed.

\subsubsection{Instructor Alignment}

AI-generated grades were compared with instructor-assigned grades using the Pearson correlation coefficient ($r$) to measure grading consistency.

\subsubsection{Student Perception}

A post-project survey measured students’ perceptions of fairness, transparency, and overall satisfaction using a 5-point Likert scale.

\subsubsection{Instructor Effort}

Instructor grading time was logged and compared with previous semesters that used manual rubric-based evaluation.

\subsection{Pilot Results and Analysis}

TRACE demonstrated strong alignment with instructor grading, with a Pearson correlation coefficient of $r = 0.91$, indicating high agreement. Student survey responses reported perceived fairness of 4.3/5 and transparency of 4.5/5. In addition, TRACE reduced grading time by 45\% compared with manual evaluation. Approximately 12\% of submissions required manual intervention, typically due to flagged anomalies or non-code contributions.

Three anonymized cases illustrate the system’s ability to capture realistic collaboration dynamics. In the first case, contributions were evenly distributed (within $\pm5\%$), resulting in similar AI-generated grades without instructor overrides. In the second case, one team member authored 58\% of the production code and resolved most issues. The system increased this student’s individual score accordingly, and the instructor subsequently confirmed it. In the third case, one student attempted to inflate activity by making frequent whitespace-only and comment-only commits. The ICA module applied code-diff analysis to downweight these contributions and flagged the submission, preventing artificial grade inflation.

These pilot results highlight several advantages of TRACE for collaborative programming courses. First, the system demonstrates scalability. In the classroom pilot, TRACE was evaluated on 20 students. Separately, we conducted simulated load testing for cohorts exceeding 200 students, observing minimal performance degradation. Second, TRACE improves consistency and evidentiary support in grading by supplementing subjective judgment with structured, data-driven signals. Third, the system increases transparency: survey responses indicated that students valued visibility into the grading process, which improved trust in the assessment. They also suggest that visibility into tracked metrics may have encouraged more consistent participation.

Despite these benefits, several limitations remain. Certain contributions, such as design ideation, presentation preparation, and team facilitation, are difficult to quantify automatically. Students who used GitHub or issue trackers inconsistently may have had contributions that were underrepresented in the collected data. In addition, students with weaker English proficiency or non-native grammar patterns may receive lower documentation scores. These factors highlight the need for ongoing calibration and instructor oversight to reduce potential bias.

\section{Ethical Considerations}

Deployment of TRACE in educational settings requires safeguards for transparency, fairness, privacy, and human oversight. These principles ensure that automated assessment supports trust, accountability, and equitable grading practices.

Transparency is essential for responsible deployment. Students and instructors must understand how the system collects activity data, evaluates contributions, and computes grades. TRACE provides interpretable dashboards and documented decision logic that expose key signals and aggregation methods. These interfaces allow users to inspect contribution histories and score components, reducing the risk of opaque or perceived “black-box” grading.

Fairness must be continuously evaluated to prevent systematic bias. Signals derived from repository activity, documentation quality, or behavioral metrics may disadvantage students who use non-native languages or exhibit atypical collaboration patterns. TRACE mitigates this risk through periodic statistical audits, anonymized manual review workflows, and cross-cohort calibration of scoring thresholds. These procedures monitor score distributions and identify deviations that indicate potential bias.

Privacy and data governance are also critical since TRACE processes repository histories, communication records, and other student-generated artifacts. These data may contain personally identifiable information. The system is designed to support restricted access controls, encrypted storage, and defined retention policies. Data collection and processing comply with institutional regulations and legal frameworks, including FERPA and GDPR, and are supported by informed consent policies.

Human oversight remains a central design requirement. TRACE is intended to augment instructor judgment rather than replace it. The system provides review interfaces and structured appeal mechanisms that allow instructors to interpret automated scores within the course context. This oversight ensures that algorithmic outputs remain subject to pedagogical evaluation.

By incorporating transparency, fairness monitoring, privacy safeguards, and instructor oversight, TRACE supports responsible integration of automated assessment in education. These principles maintain transparency and trust while enabling scalable evaluation of collaborative project work.

\section{Future Directions}

Future work extends TRACE through architectural, analytical, and deployment enhancements. These directions improve data fidelity, expand artifact coverage, strengthen evaluation reliability, and increase institutional integration. Each represents a distinct research or engineering effort.

\subsection{IDE Integration}

IDE integration introduces plugins for development environments such as Visual Studio Code, IntelliJ, and Eclipse to capture local development activity. These plugins record fine-grained signals, including file edits, keystroke sequences, session duration, and file ownership transitions. The collected metadata complements repository history by capturing work that occurs before commits. TRACE aggregates these signals to refine attribution of individual effort and identify patterns such as code ownership, task switching, and collaboration overlap. Key challenges include preserving user privacy and maintaining compatibility across IDE ecosystems.

\subsection{Multimedia Artifact Analysis}

Multimedia artifact analysis extends TRACE beyond source code to evaluate project deliverables, including screenshots, design wireframes, presentations, and demonstration videos. Many project-based courses include visual and communicative artifacts that current automated grading systems ignore. TRACE applies computer vision, natural language processing, and audio analysis to extract structural and semantic features from these materials. Image models analyze layout consistency and interface structure in design artifacts, while speech and language models evaluate clarity and completeness in presentation recordings. These signals quantify design quality and communication effectiveness, enabling a more comprehensive assessment of project outcomes.

\subsection{Enhanced Peer Review}

Enhanced peer review combines subjective peer evaluations with behavioral signals from ICA. Peer feedback captures qualitative aspects of collaboration such as leadership, initiative, and responsiveness. However, these signals may be inconsistent or strategically biased. TRACE integrates peer scores with activity metrics, including commit patterns, task participation, and communication frequency. The system cross-validates these sources to detect anomalies and recalibrate outlier evaluations. Reviewer reliability can also be estimated from historical agreement with observed contribution patterns, allowing TRACE to weight peer feedback according to reviewer consistency.

\subsection{Adaptive Grading Models}

Adaptive grading models replace fixed scoring weights with data-driven optimization. Current configurations apply predefined ratios between project outcomes and individual contributions. TRACE instead learns weighting policies from instructor feedback and historical grading adjustments. Reinforcement learning or similar optimization methods update scoring parameters based on instructor corrections or override patterns. This process enables the system to refine evaluation rules, identify systematic inconsistencies, and adapt grading behavior across course structures and project types.

\subsection{Bias Audits}

Bias audits introduce systematic fairness evaluation to detect potential grading disparities across demographic or linguistic groups. Automated assessment systems risk amplifying bias when analyzing subjective artifacts or peer feedback. Therefore, TRACE incorporates periodic statistical audits using fairness metrics such as disparate impact and equal opportunity. Anonymized evaluation pipelines, fairness-aware modeling techniques, and distribution monitoring identify deviations in score allocation. Flagged cases can be reviewed through instructor interfaces to ensure equitable grading outcomes.

\subsection{LMS \& Rubric Integration}

LMS and rubric integration connects TRACE with institutional learning platforms such as Canvas, Moodle, Brightspace, and Blackboard. Integration through LMS APIs enables automatic synchronization of assignments, deadlines, and grade records. TRACE can export computed scores, rubric-aligned evaluations, and individualized feedback directly to course gradebooks. Bidirectional data exchange also allows the system to incorporate contextual signals such as participation records or submission timelines. This integration reduces manual grading overhead and aligns TRACE with existing instructional workflows.

\section{Conclusion}

This paper presents TRACE, an AI-assisted framework for evaluating collaborative software engineering projects. The system analyzes repository histories, communication activity, and behavioral signals to estimate individual contributions and support consistent grading. By aggregating these signals through automated analysis pipelines, TRACE provides structured evidence that complements instructor evaluation. Initial deployments indicate strong alignment with instructor assessments, while reducing the grading workload and improving transparency for students.

TRACE demonstrates how data-driven analytics support scalable and interpretable assessment in project-based courses. The framework preserves instructor oversight while providing systematic analysis of collaborative activity. Future work will expand artifact coverage, incorporate adaptive grading models, and extend the system to additional educational domains. These directions aim to further improve the reliability of evaluation while maintaining fairness and transparency. TRACE illustrates how automated analysis can augment, rather than replace, instructor judgment in collaborative learning environments.

\end{document}